\documentclass[twocolumn]{aastex63}

\usepackage[T1]{fontenc}
\usepackage{ae,aecompl}

\usepackage{graphicx}	
\usepackage{amsmath}	
\usepackage{amssymb}	
\usepackage{url}
\usepackage{xspace}
\usepackage{multirow}
\usepackage{lineno}

\def\arcdeg{\hbox{$^\circ$}}

\def\arcsec{\hbox{$^{\prime\prime}$}}

\def\nustar{\textit{NuSTAR}\xspace}

\def\nicer{\textit{NICER}\xspace}

\def\rxte{\textit{RXTE}\xspace}
\def\srg{\textit{SRG}\xspace}
\def\ero{{eROSITA}}
\def\maxi{\textit{MAXI}\xspace}

\def\src{Swift\,J1626.6-5156\xspace}

\newcommand {\be}{\begin {equation}}
\newcommand {\ee}{\end {equation}}

\def\a{$^{\mbox{\small a}}$}
\def\b{$^{\mbox{\small b}}$}
\def\c{$^{\mbox{\small c}}$}
\def\d{$^{\mbox{\small d}}$}
\def\e{$^{\mbox{\small e}}$}

\received{XXX, 2021}
\revised{XXX, 2021}
\accepted{17 June, 2021}
\submitjournal{ApJL}

\shorttitle{Cyclotron line absorption features in \src}
\shortauthors{Molkov et al.}

\begin{document}

\title{Discovery of the 5 keV cyclotron line followed by three harmonics in \src}

\correspondingauthor{Sergey Molkov}
\email{molkov@iki.rssi.ru}

\author{S. Molkov}
\affiliation{Space Research Institute, Russian Academy of Sciences,
Profsoyuznaya 84/32, 117997 Moscow, Russia}
\affiliation{Moscow Institute of Physics and Technology, Moscow region,
141701 Dolgoprudnyi, Russia}

\author{V. Doroshenko}
\affiliation{Institute for Astronomy and Astrophysics, University of
T\"ubingen, Sand 1, 72026 T\"ubingen, Germany}

\author{A. Lutovinov}
\affiliation{Space Research Institute,
 Russian Academy of Sciences,
Profsoyuznaya 84/32, 117997 Moscow, Russia}
\affiliation{Moscow Institute of Physics and Technology, Moscow region,
141701 Dolgoprudnyi, Russia}

\author{S. Tsygankov}
\affiliation{Department of Physics and Astronomy,
 FI-20014 University of Turku, Finland}
\affiliation{Space Research Institute,
 Russian Academy of Sciences,
Profsoyuznaya 84/32, 117997 Moscow, Russia}

\author{A. Santangelo}
\affiliation{Institute for Astronomy and Astrophysics, University of
T\"ubingen, Sand 1, 72026 T\"ubingen, Germany}

\author{I. Mereminskiy}
\affiliation{Space Research Institute,
 Russian Academy of Sciences,
Profsoyuznaya 84/32, 117997 Moscow, Russia}

\author{A. Semena}
\affiliation{Space Research Institute,
 Russian Academy of Sciences,
Profsoyuznaya 84/32, 117997 Moscow, Russia}

\begin{abstract}
We report on observations of the Be/X-ray binary system \src performed with \nustar during a short
outburst in March 2021, following its detection of by the \maxi\ monitor and
Spektrum-Roentgen-Gamma (\srg) observatory. Our analysis of the broadband X-ray spectrum of
the source confirms the presence of two absorption-like features at energies $E\sim9$ and
$E\sim17$ keV previously reported in literature and interpreted as the fundamental cyclotron
resonance scattering feature (CRSF) and its first harmonic (based on \rxte data). The better
sensitivity and energy resolution of \nustar, combined with the low energy coverage of \nicer,
allowed us to detect two additional absorption-like features at $E\sim4.9$\,keV and $E\sim13$ keV.
We conclude, therefore, that in total four cyclotron lines are observed in the spectrum of \src:
the fundamental CRSF at $E\sim4.9$\,keV and three higher spaced harmonics. This discovery
makes \src the second accreting pulsar, after 4U\,0115+63, whose spectrum is characterized by more
than three lines of a cyclotronic origin, and implies the source has the weakest confirmed magnetic
field among all X-ray pulsars $B\sim4\times10^{11}$\,G. This discovery makes \src\ one of prime
targets for the upcoming X-ray polarimetry missions covering soft X-ray band such as IXPE and eXTP.

\end{abstract}




\section{Introduction} \label{sec:intro}

\src is a transient X-ray pulsar (XRP) with a spin period of $\sim15$\,s
discovered on December 18, 2005 by the Swift Burst Alert Telescope
(BAT) during a giant outburst \citep[][]{2005GCN..4361....1K}. The outburst lasted for approximately
half a year and was followed by a longer period of a few years during which the object remained active
and strongly variable on timescales from 45 to 95 days \citep[][]{2008A&A...485..797R,2010ApJ...711.1306B}.

The most complete study of the system to date
has been conducted by \citet[][]{2011A&A...533A..23R}, who based on multi-frequency observations
concluded that \src is a Be/X-ray binary (BeXRB) with a B0Ve companion located at a distance of
$D\sim 10$ kpc. This estimate is rather uncertain and also the Gaia EDR3 estimates 
are in the range of 5.8-12\,kpc \citep{2021AJ....161..147B}. Here we adopt 10\,kpc distance for
easier comparison with previous results. 
The counterpart (2MASS16263652-5156305) shows strong $H{\alpha}$ emission 
\citep[][]{2006ATel..831....1N}, typical of a Be star.
The $\sim 15.3$\,s spin period of the neutron star \citep[][]{2005ATel..678....1P,2005ATel..679....1M}
and the $132.9$\,days orbital binary period \citep[][]{2010ApJ...711.1306B} are also typical for
Be-systems \citep[see e.g.][]{1986MNRAS.220.1047C}. At the same time, the binary orbit is near circular
unlike most other BeXRBs. The optical counterpart is also rather faint in the infrared for a Be
star \citep[][]{2006ATel..713....1R}. Finally, the observed outburst light curve is also not typical
for Be/X-ray binaries, so the system is not without peculiarities. 

In the X-ray band the source was extensively studied using RXTE observations. The main result was
a significant detection in the energy spectra of two cyclotron resonance scattering features (CRSFs)
at $E\sim10$\,keV and $E\sim18$\,keV \citep[][]{2013ApJ...762...61D}, which implied a magnetic field
of $\sim10^{12}$\,G in the line forming region.

\begin{figure}[!t]
\vspace{-3cm}
\hspace{-1cm}\includegraphics[width=1.2\columnwidth]{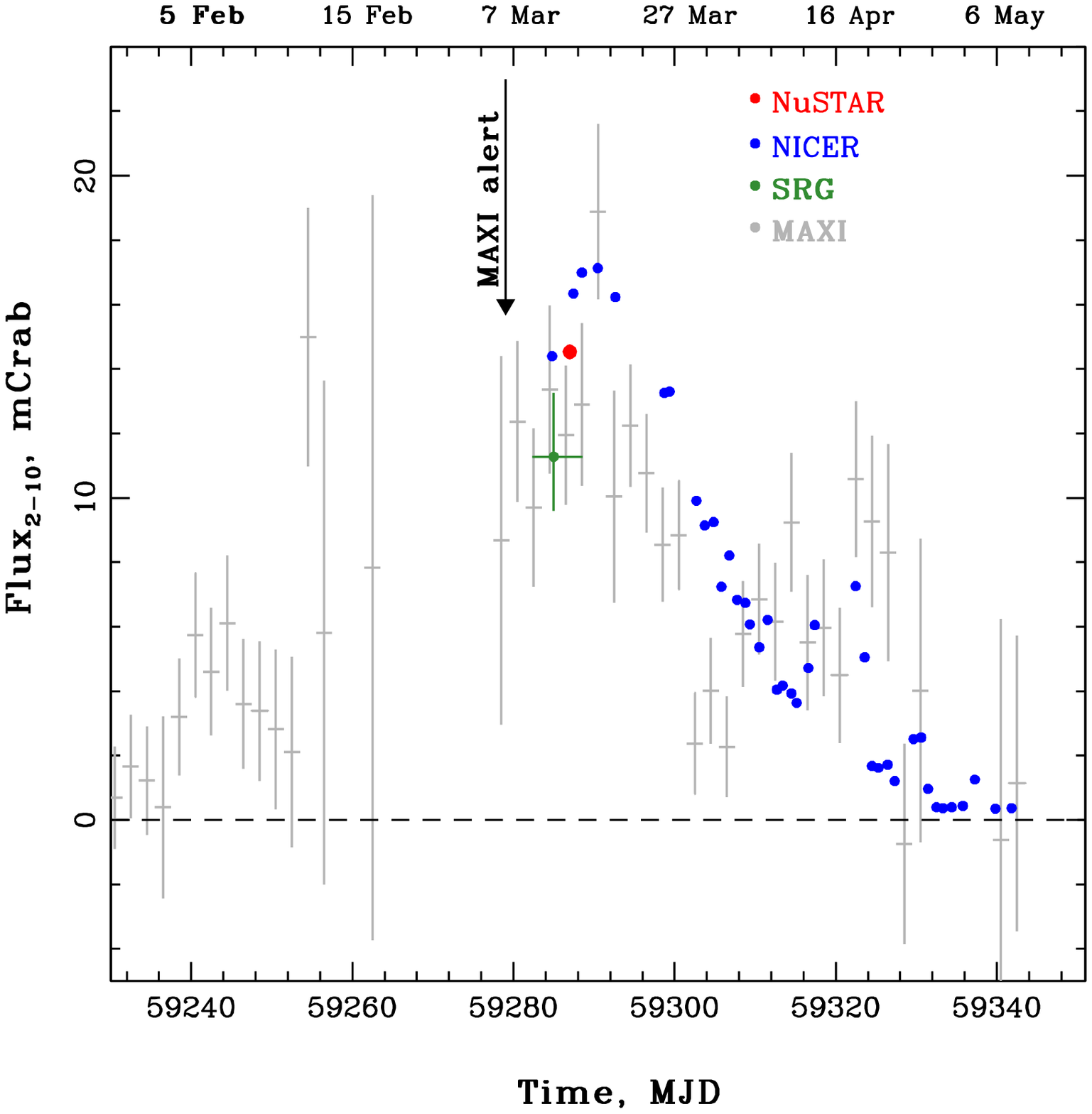}
\vspace{-3cm}
\caption{The {\it MAXI } light curve of \src around the 2021
  outburst (grey crosses). On top of this curve fluxes obtained from
 the {\it NuSTAR} (red point) and NICER (blue points) data
  are presented. All fluxes are given in mCrab units in the $2-10$ keV energy band. 
  Date of the MAXI outburst trigger is shown with vertical arrow.}
\label{fig:outb_hist}
\end{figure}

In 2008 the pulsar went into a low state characterised by a lowest observed luminosity
of $\sim(3-4)\times10^{33}$\,erg\,s$^{-1}$ in the 0.5-10 keV energy band \citep[][]{2017MNRAS.470..126T} 
and remained undetected by all-sky monitors until 2021. 
On March 2021, \maxi/GSC significantly detected the source \citep[][]{2021ATel14454....1N},
suggesting that \src had started a new
giant outburst after fifteen years of relative quiescence (Fig.~\ref{fig:outb_hist}).
However, in the following few weeks, no giant outburst developed and only a modest flux enhancement
was observed \citep{2021ATel14462....1M}. Moreover, the analysis of archival \maxi\ data revealed that
the source actually shows flaring activity from time to time \citep[][]{2021ATel14498....1N}, i.e.
confirming likely accretion in quiescence \citep{2017MNRAS.470..126T}.

Here we report results of observations of \src\ performed in March 2021 with
the \srg, \textit{NuSTAR}, and \nicer\ missions covering a broad energy range from 0.2 to 78\,keV.

\section{Observations and data reduction}

As already mentioned, \src remained in a state of relative quiescence until 2021, after the end of the
giant outburst and following activity in 2005-2008. First evidence for a renewed activity of the source
was reported by \citep[][]{2021ATel14454....1N} using \maxi\ \citep[][]{2009PASJ...61..999M} data. Initially,
it was proposed that the source was entering into a new giant outburst, but the follow-up monitoring
campaigns revealed that this was not the case.

To follow the flux evolution of the source in this flare, we used publicly available data
from \maxi\footnote{\url{http://maxi.riken.jp/star_data/J1626-519/J1626-519.html}.
Data are multiplied by 2 because of about half the sky region to obtain the fluxes is
masked to avoid count leaks from a nearby source \citep[][]{2021ATel14454....1N}}.
The resulting light curve re-binned to 2 day time intervals is shown on Fig.~\ref{fig:outb_hist}, where
all fluxes are given in mCrab units in the $2-10$ keV range.
The source flux measured with the Neutron star Interior Composition Explorer
\citep[NICER,][]{2017NatAs...1..895G} is plotted in the same figure. The public data were downloaded
from the HEASARC archive system and processed with {\sc heasoft} v.6.28, using the NICER Calibration
Database (CALDB) version 20200722. For background estimation we used the {\sc nibackgen3C50} tool. 

As shown in Fig.~\ref{fig:outb_hist}, \nicer performed many observations covering a large part of the
outburst of 2021. Since in this article we focus on the broadband spectral analysis, we only report
the analysis of the first observation (ID:4202070101) performed on Mar. 11, 2021, temporally close
and at similar flux level of the \nustar observation (see below).

We used data obtained by the \srg\ observatory \citep{sunyaev21} during the third all sky survey to
assess source flux at early stages of the outburst and improve the low-energy coverage for
phase-averaged spectra (Fig.~\ref{fig:outb_hist}). The sky region around \src was scanned
by \srg on Mar 12, 2021. Both the {\it Mikhail Pavlinsky} ART-XC
telescope \citep[][]{2021arXiv210312479P} and the \ero\ telescope \citep[][]{2021A&A...647A...1P} onboard
the \srg\ observatory detected the source with the high significance \citep{2021ATel14462....1M}.
The source flux in the 2-10 keV energy band resulted from the joint fit of the ART-XC and \ero\ spectra
is shown in Fig.~\ref{fig:outb_hist}.

\begin{figure}[!t]
\vspace{-2cm}
\hspace{-0.8cm}\includegraphics[width=1.2\columnwidth]{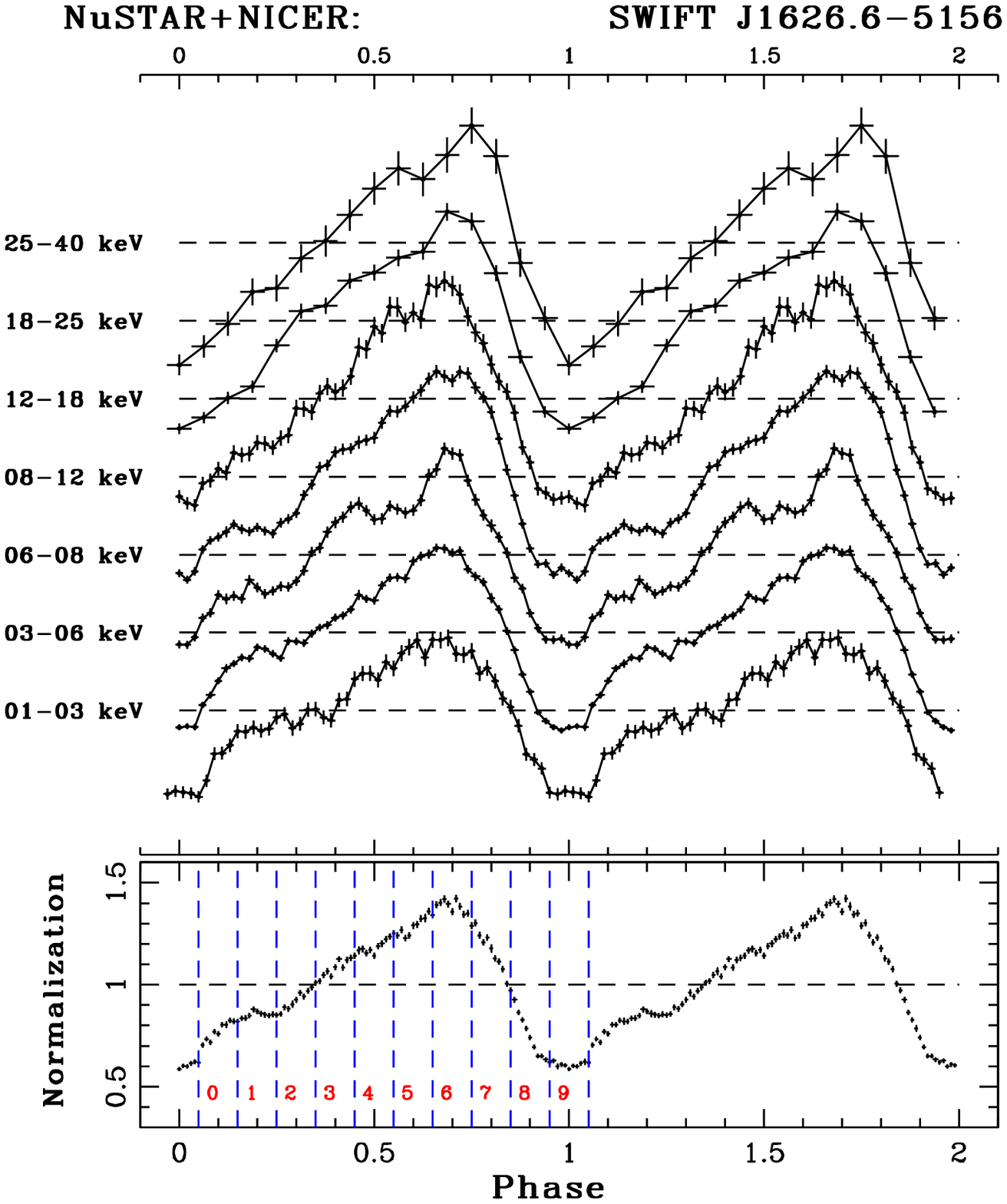}
\vspace{-3cm}
\caption{Energy-resolved pulse profiles of \src obtained with {\it NuSTAR} (above 3 keV)
and NICER (1-3 keV).
On the bottom panel an averaged NuSTAR pulse profile in the 3-50 keV energy band
is shown. Vertical lines define boundaries of phase
bins selected for spectral analysis.
}
\label{fig:fold_crv}
\end{figure}

Based on the above data we requested follow-up observations with the {\it NuSTAR} (Nuclear Spectroscopic
Telescope ARray) observatory. It consists of two X-ray telescope modules, to which we refer to as FPMA
and FPMB \citep{2013ApJ...770..103H}. It provides X-ray imaging, spectroscopy and timing in the energy
range of 3-79\,keV with an angular resolution of 18\arcsec\ (FWHM) and spectral
resolution of 400\,eV (FWHM) at 10\,keV. {\it NuSTAR} performed one observation of \src on Mar 13-14, 2021,
near the peak of the flare  (ObsIDs:90701311002) with the on-source
exposure of $\sim56$\,ks (see Fig.~\ref{fig:outb_hist}). {\it NuSTAR} data were processed with the
standard {\it  NuSTAR} Data Analysis Software ({\sc nustardas\_19Jun20\_v2.0.0}) provided
under {\sc heasoft} v6.28
with {\sc caldb} version 20201217.

All {\it NuSTAR} and \nicer spectra were grouped to have at least 25 counts per bin and at least 3 detector
channels, to ensure that the binning of the spectra matches the energy resolution of the detectors. 
The final data analysis (timing and spectral) was performed with the {\sc heasoft 6.28} software package.
All uncertainties are quoted at the {$1\sigma$} confidence level, if not stated otherwise.

\section{Results}
In this section we present the detailed results of spectral (including pulse phase-resolved) and
timing analysis of {\it NuSTAR} and \nicer data.

\subsection{Energy-resolved pulse profile}

The orbital ephemerides for \src\ are not well known \citep[see][and references therein]{2011MNRAS.415.1523I}.
This is not relevant for the current work as the duration of the \nustar observation is much shorter
than the expected orbital period. Therefore only barycentric correction was applied to the light curves
and the pulse period of $P=15.33962(1)$~s used for phase-resolved spectroscopy was determined.
Uncertainty for the pulse period value
was calculated from the Monte-Carlo simulations \citep[][]{boldin13}.  

Energy resolved pulse profiles obtained with {\it NuSTAR} in the $3-40$ keV energy interval and
with \nicer in the $1-3$ keV energy interval, folded with the aforementioned period are
presented in Fig.~\ref{fig:fold_crv}. Phase '0' corresponds to the minimum of the light
curve folded in the whole \nustar energy band. 
The source pulse profile is mainly characterized by one rather broad peak and demonstrates
some evolution of both the shape and pulsed amplitude with the phase. In particular
in softer energies, the profile shows two sub-peaks near phases 0.1 and 0.4. As the energy
increases the features disappear.

\begin{figure}[!th]
\vspace{-2.5cm}
\hspace{-0.5cm}\includegraphics[width=1.1\columnwidth]{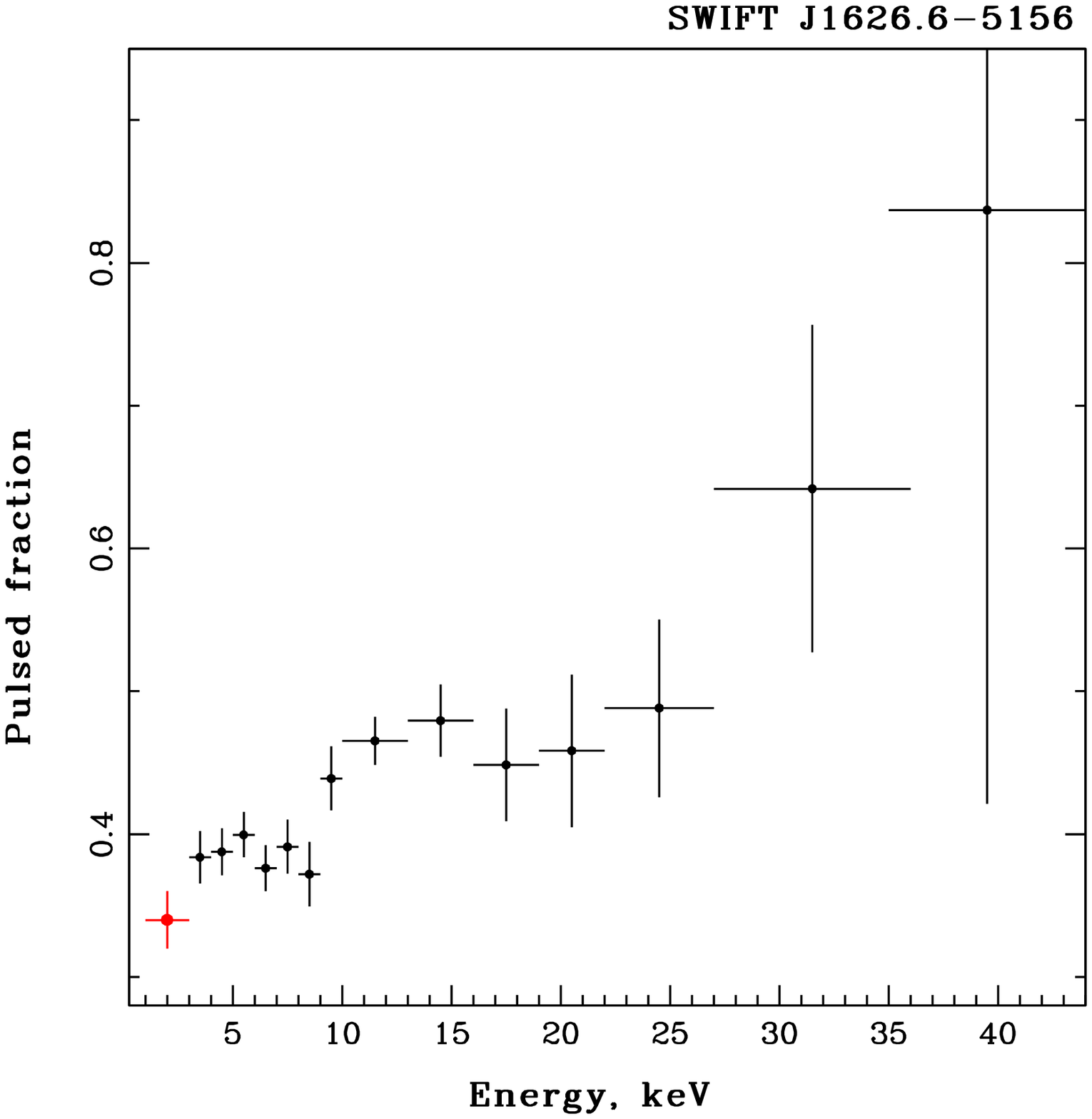}
\vspace{-3cm}
\caption{Dependence of the pulsed fraction of \src on the energy. NuSTAR values shown by
black color, NICER 1-3 keV measurement -- in red.
}
\label{fig:puls_frac}
\end{figure}

The pulsed fraction gradually increases with the energy from $\sim 40\%$ at 
$3-5$ keV to $\sim 60\%$ at $30-40$ keV (Fig.~\ref{fig:puls_frac}).
Such a behaviour is typical for the majority of bright XRPs \citep[see,
e.g., ][]{2009AstL...35..433L}. Furthermore, a sharp decrease of the pulsed fraction is
observed around 20\,keV, that roughly corresponds to the energy of the cyclotron line harmonic
reported by \cite{2013ApJ...762...61D}. 
Hints of decrease in the pulse fraction are also observed around $\sim10$\,keV, reported
by those authors as a fundamental energy of cyclotron line, and at even lower energies,
i.e., below 10\,keV. The counting statistics do not allow making any significant
conclusions.
It is worthy to note that very similar decrease of the pulsed fraction near the first
harmonic of the cyclotron line was early found by \citet{2009A&A...498..825F} in 4U\,0115+63. Both
these findings are quite rare as an increase of the pulsed fraction is usually observed near
the cyclotron line and its harmonics
\citep[see, e.g.,][]{2007AstL...33..368T,2009AstL...35..433L,2019MNRAS.482L..14S}.      

\subsection{Phase-averaged spectrum}

\begin{figure}[!th]
\vspace{-2cm}
\hspace{-3mm}\includegraphics[width=1.1\columnwidth]{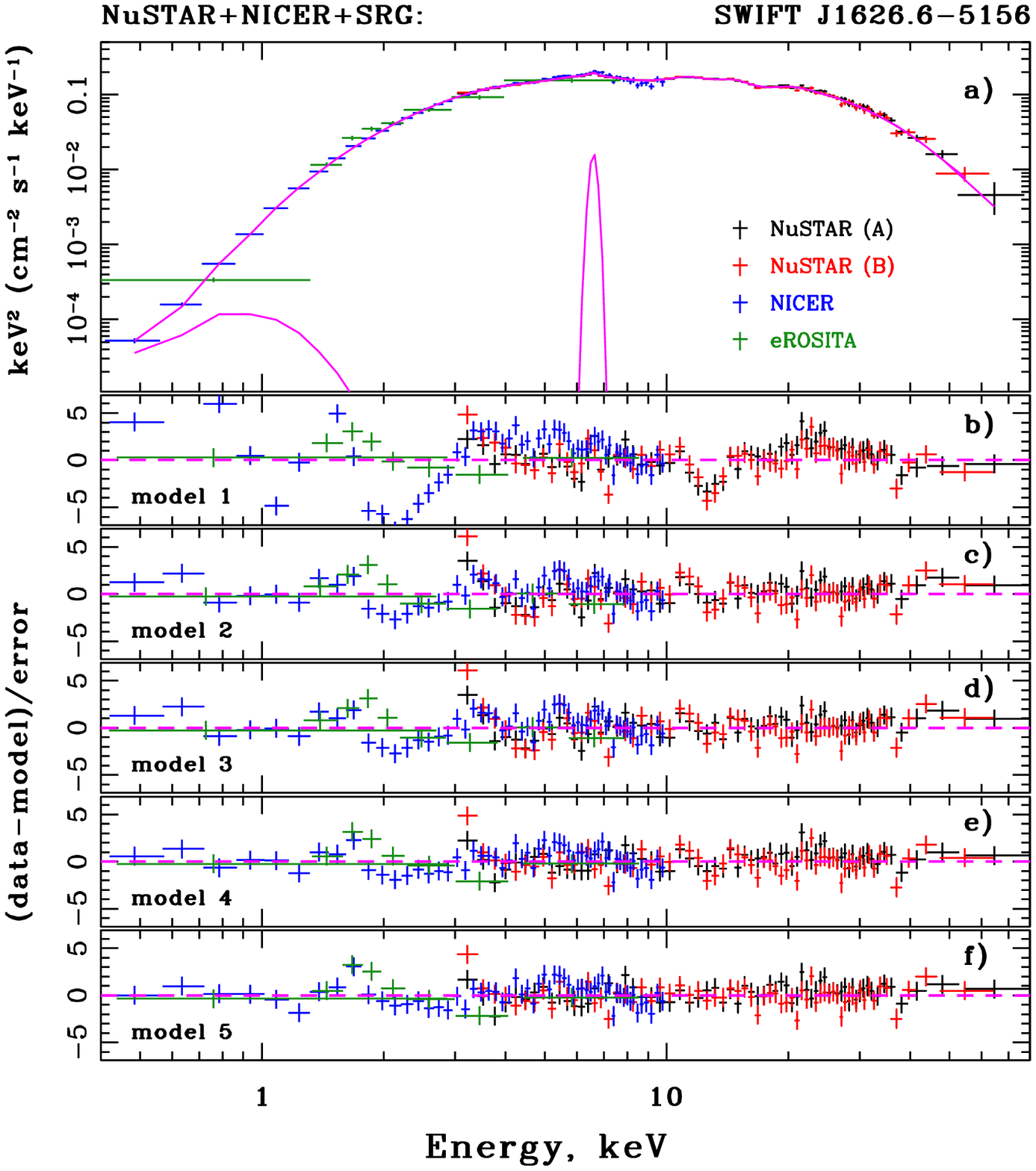}
\vspace{-2.7cm}
\caption{The phase-averaged energy spectrum of \src reconstructed in
a wide energy range with the {\it NuSTAR, NICER} and {\it SRG}/\ero\ instruments (upper panel). 
The five bottom panels show residuals for the five spectral models
(see the text and Table~ \ref{tab:spe_mean}).}
\label{fig:mean_spec_all}
\end{figure}

The spectrum of \src is typical for accreting XRPs  \citep[see,
e.g.,][]{1989PASJ...41....1N, 2005AstL...31..729F}. It is characterized by an
exponential cutoff at high energies (Fig.\,\ref{fig:mean_spec_all}), that 
can be explained in terms of the Comptonization processes in hot emission
regions \citep[see, e.g., ][]{1980A&A....86..121S, 1985ApJ...299..138M, 1994ApJ...434..570T}.
We modelled the broadband continuum spectrum with two commonly used
phenomenological models: a power law with an exponential cutoff (\texttt{cutoffpl} in the {\sc
xspec} package, hereafter \texttt{model1}) and a thermal Comptonization model (\texttt{comptt},
hereafter  \texttt{model2}). To take into account the uncertainty in calibrations of two
modules of {\it NuSTAR} a cross-calibration constant $C_{modB}$ was included in
all spectral fits. Furthermore, two cross-calibration constants were added for
\nicer ($C_{NIC}$), and \ero ($C_{eRo}$) spectra, to compensate for some flux difference between
the observations by these instruments (we assume that the spectrum shape does not change
significantly). Depending on the continuum model, the inclusion of a soft black body component
with the temperature of $\sim0.1$\,keV improves the fit quality in the softer part of the
spectrum \citep[see also, ][]{2021ATel14457....1I}. We included in the fit a Gaussian function
to model an emission of the neutral fluorescence iron line (\texttt{gauss}), and
the \texttt{phabs} component to take into account interstellar absorption. For both continua,
an inclusion of two earlier reported absorption features around 10 and 20\,keV was also
necessary to obtain a meaningful fit (\texttt{gabs}). Results of the fit are
presented in Fig.\,\ref{fig:mean_spec_all}a and Table.~\ref{tab:spe_mean}.

The quality of the fit is however not acceptable for both models. 
Although the reduced $\chi^2$ value is around $\sim1.2$ for \texttt{model2}, an
assessment of the fit quality using simulations (by means of a \texttt{goodness}
command in \texttt{Xspec}) revealed an unsatisfactory quality of the
fit (Table~\ref{tab:spe_mean}). Poor quality of the fit is also revealed by residuals which
are observed around 5 and 13\,keV (Fig.\,\ref{fig:mean_spec_all}b,c). The quality of the fit
could be improved by including additional absorption line features near
13 keV (\texttt{model3}, Fig.\,\ref{fig:mean_spec_all}d) or near
5 keV (\texttt{model4}, Fig.\,\ref{fig:mean_spec_all}e).
We note, however, that including only one of the two lines does not result in a statistically
acceptable fit. If we add only a 13 keV line then neither fit-statistic nor Bayesian information
criteria (appropriate for comparison
of non-nested models) indicate significant improvement and the goodness parameter
remains at an unacceptable level. Adding only the 5\,keV absorption feature
leads to better approximation and goodness parameter becomes closer to 50\%. But only
the inclusion of both lines (\texttt{model5}) dramatically
improves the quality of the fit (Fig.\,\ref{fig:mean_spec_all}f) and makes it adequate both
in terms of fit-statistics and Bayesian criteria, and as assessed with simulations
using \texttt{goodness} command. 

We finally conclude, therefore, that a statistically acceptable fit of the averaged spectrum
can only be obtained if all four absorption features are included in the model. This, together
with the fact that the centroid energies of these features appear to be harmonically spaced for
a fundamental line energy at $E\sim5$ keV, strongly indicates that all four features
have a physical origin.

On the other hand, also the best-fit discussed above has some issues.
As it can be seen in Table~\ref{tab:spe_mean}, the centroid of the iron line energy becomes
significantly higher than the expected value of $6.4$ keV in \texttt{model4} and \texttt{model5}.
We note that the equivalent width of this line is quite low, which might imply that the available
statistics actually does not allow to significantly detect the line and constrain its parameters,
so the increase of its apparent energy might be an artifact of the fit. Indeed, the shifted energy
of the iron line appears in models with the addition of the soft absorption feature
at $E\sim5$ keV (\texttt{model4} and \texttt{model5}). Taking into account that the both models
contain an absorption line around $E\sim9$ keV, one can imagine the appearance of the emission-like
feature in the region around 6-7 keV, especially if the centroid energies of these absorption
features vary with the pulse phase. To investigate this issue in detail, we performed also the
phase-resolved spectral analysis.

\begin{table*}[!h]
\small
\begin{center}
\caption{Best-fitting results for the \src averaged-spectrum for five models}
\label{tab:spe_mean}
\begin{tabular}{l|c|c|c|c|c}
\hline
\hline
$~~~$ & {\small\bf model1} & {\small\bf model2} & {\small\bf model3} & {\small\bf model4} & {\small\bf model5}\\
$~~~$ & {\small\bf pha$\times$} & {\small\bf  pha$\times$} & {\small\bf  pha$\times$} & {\small\bf  pha$\times$} & {\small\bf  pha$\times$}\\
$~~~$ & {\small\bf (bb+cutoff+ga)} & {\small\bf (bb+comptt+ga)} & {\small\bf (bb+comptt+ga)} & {\small\bf (bb+comptt+ga)} & {\small\bf (bb+comptt+ga)}\\
$~~~$ & {\small\bf $\rm \times2$~gabs} & {\small\bf $\rm \times2$~gabs} & {\small\bf $\rm\times3$~gabs} & {\small\bf $\rm \times3$~gabs} & {\small\bf $\rm\times4$~gabs}\\
\hline
Parameter  & Value  & Value & Value  & Value & Value\\
\hline
$N_{\rm H}$~$^a$           &  $2.23\pm0.02$ &  $0.77\pm0.03$  &  $0.78\pm0.03$  &  $0.61\pm0.04$ & $0.42\pm0.05$  \\ 
\hline
$kT_{\rm BB}$, keV         &  $0.08\pm0.01$ &  $0.14\pm0.01$   &  $0.14\pm0.01$  &  $0.14\pm0.01$    &  $0.13\pm0.03$  \\
$A_{\rm BB}~^b, \times10^{3}$  &   $654\pm92$ & $0.33\pm0.06$   &  $0.34\pm0.07$  &  $0.27\pm0.03$    &  $0.03\pm0.01$  \\
\hline
$\Gamma$                   &  $1.14\pm0.01$ &  $-$  & $-$  &  $-$   &  $-$ \\
$E_{\rm fold}$, keV        &  $10.12\pm0.12$ &  $-$  & $-$  &  $-$    &  $-$\\
$A_{\rm cut}~^c$, $\times10^{1}$  &  $0.67\pm0.01$ &  $-$   & $-$   &  $-$   &  $-$ \\
\hline
$T_0$, keV                 &   $-$ & $0.89\pm0.01$   &  $0.89\pm0.01$  &  $0.99\pm0.02$    &  $1.17\pm0.08$  \\
$T_{\rm p}$, keV           &   $-$ & $5.36\pm0.04$   &  $5.32\pm0.04$  &  $5.65\pm0.07$    &  $5.74\pm0.16$  \\
$\tau_{\rm p}$             &   $-$ & $4.22\pm0.04$   &  $4.26\pm0.04$  &  $3.83\pm0.08$    &  $3.43\pm0.25$  \\
$A_{\rm comp}$,~$\times10^{1}$ &   $-$ & $0.156\pm0.002$ &  $0.157\pm0.002$&  $0.148\pm0.002$  &  $0.157\pm0.005$\\
\hline
$E_{\rm Fe}$, keV          &   $6.38\pm0.05$ & $6.41\pm0.03$   &  $6.41\pm0.03$  &  $6.57\pm0.03$    &  $6.59\pm0.03$  \\ 
$\sigma_{\rm Fe}$, keV     &   $0.61\pm0.05$ & $0.46\pm0.03$   &  $0.46\pm0.03$  &  $0.16\pm0.05$    &  $0.13\pm0.05$  \\ 
$A_{\rm Fe}~^d,~\times10^{3}$   &   $0.64\pm0.07$ & $0.58\pm0.05$   &  $0.58\pm0.05$  &  $0.18\pm0.03$    &  $0.16\pm0.04$  \\
$EW_{\rm Fe}$, eV         &   $146$ & $142$           &  $141$          &  $42$             &  $32$           \\ 
\hline
$E_{cyc1}$, keV            &  $-$ &  $-$             &  $-$            &  $4.73\pm0.05$    &  $4.82\pm0.05$  \\
$\sigma_{cyc1}$, keV       &   $-$ & $-$             &  $-$            &  $0.58\pm0.09$    &  $0.93\pm0.09$  \\
$\tau_{cyc1}$, keV         &   $-$ & $-$             &  $-$            &  $0.12\pm0.03$    &  $0.47\pm0.14$  \\
\hline
$E_{cyc2}$, keV            &   $9.02\pm0.06$ & $8.94\pm0.04$   &  $8.95\pm0.05$  &  $8.78\pm0.05$    &  $8.63\pm0.06$  \\
$\sigma_{cyc2}$, keV       &   $1.18\pm0.08$ & $0.77\pm0.06$   &  $0.79\pm0.06$  &  $1.00\pm0.07$    &  $1.74\pm0.20$  \\
$\tau_{cyc2}$, keV         &   $0.45\pm0.03$ & $0.21\pm0.02$   &  $0.22\pm0.02$  &  $0.38\pm0.05$    &  $1.51\pm0.56$  \\
\hline
$E_{cyc3}$, keV            &   $-$ & $-$             &  $12.96\pm0.20$ &  $-$              &  $12.84\pm0.11$ \\
$\sigma_{cyc3}$, keV       &   $-$ & $-$             &  $0.57\pm0.26$  &  $-$              &  $0.97\pm0.15$  \\
$\tau_{cyc3}$, keV         &   $-$ & $-$             &  $0.06\pm0.03$  &  $-$              &  $0.30\pm0.11$  \\
\hline
$E_{cyc4}$, keV            &   $17.00\pm0.11$ & $17.33\pm0.10$  &  $17.37\pm0.10$ &  $17.21\pm0.10$   &  $17.09\pm0.14$ \\
$\sigma_{cyc4}$, keV       &   $1.06\pm0.11$ & $1.28\pm0.11$   &  $1.45\pm0.13$  &  $1.02\pm0.11$    &  $1.65\pm0.21$  \\
$\tau_{cyc4}$, keV         &   $0.37\pm0.04$ & $0.52\pm0.05$   &  $0.63\pm0.07$  &  $0.38\pm0.04$    &  $0.82\pm0.18$  \\
\hline
$C_{\rm modB}$                &   $1.064\pm0.003$ & $1.064\pm0.003$ &  $1.064\pm0.003$ &  $1.064\pm0.003$ &  $1.064\pm0.005$\\ 
$C_{\rm NIC}$                &   $0.917\pm0.005$ & $0.944\pm0.005$ &  $0.944\pm0.005$ &  $0.946\pm0.005$ &  $0.948\pm0.005$\\ 
$C_{\rm eRo}$                &   $0.772\pm0.003$ & $0.802\pm0.003$ &  $0.802\pm0.003$ &  $0.804\pm0.003$ &  $0.805\pm0.003$\\ 
$F_{\rm X}~^e,~\times10^{10}$   &   $5.5$ & $5.5$      &  $5.5$           &  $5.5$           &  $5.5$          \\ 
\hline
$\chi^2$ (d.o.f.)          &  1604.0 (864)  & 1027.6 (863)  &  1015.4 (860)  &  930.3 (860)  &  906.9 (857)    \\
goodness                   &  100\%   & 99\%            &   99\%           &  74\%            &   54\%          \\
BIC                        &  $1657$   &   $1084$     &   $1080$         &    $995$         &     $981$     \\
$\Delta$~BIC               &   $-$ &  $-$   &$BIC_{model2}-$& $BIC_{model2}-$&  $BIC_{model4}-$ \\
$ $               &   $ $ &  $ $   &$BIC_{model3}=4$& $BIC_{model4}=89$&  $BIC_{model5}=14$ \\
\hline
\end{tabular}
\end{center}

\begin{tabular}{l l}
\small
{\bf Notes.} & Here $N_H$ is the column density, $kT_{BB}$ is the black-body temperature, $\Gamma$ is the power-law photon index,\\
{~} & $E_{\rm fold}$ is the folding energy of the cutoff power law. $T_0$, $T_{\rm p}$ and $\tau_{\rm p}$ are the seed photons temperature,\\
{~} & the plasma temperature and the plasma optical depth Comptonization model parameters. $E_{\rm Fe}$, $\sigma_{\rm Fe}$ and $EW_{\rm Fe}$ are\\
{~} & the iron line energy, width and equivalent width, respectively. $E_{cyc}$, $\sigma_{cyc}$ and $\tau_{cyc}$ are the energy,\\
{~} & width and optical depth of cyclotron lines.\\
$^a$ & Value of $N_H$ is in units of $10^{22}$ atom cm$^{-2}$\\
$^b$ & Normalization parameter calculated as $L_{39}/D_{10}^2$, where $L_{39}$ is the source luminosity in units of $10^{39}$ erg/s and $D_{10}$\\
$~$ & is the distance to the source in units of 10 kpc\\
$^c$ & Units are {\it photons $keV^{-1}$ cm$^{-2}$ $s^{-1}$} at 1 keV\\
$^d$ & Total {\it photons cm$^{-2}$ $s^{-1}$} in the line\\
$^e$ & Model flux in the 3-50 keV energy band in units of {\it erg cm$^{-2}$ $s^{-1}$}.

\end{tabular}

\end{table*}

\begin{figure}[!t]
\vspace{-2.0cm}
\hspace{-0.5cm}\includegraphics[width=1.1\columnwidth]{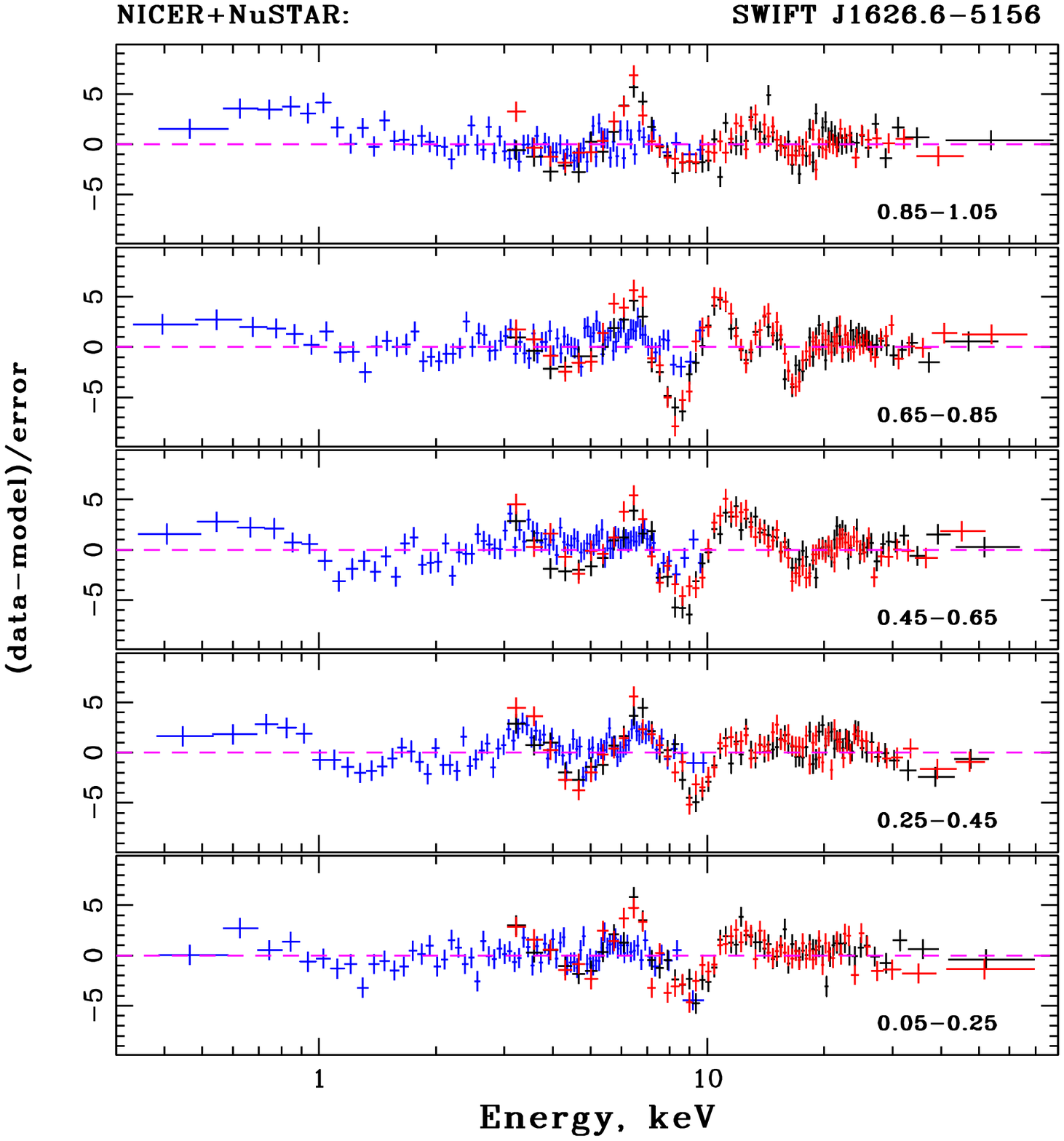}
\vspace{-2.5cm}
\caption{
Residuals of the pulse-phase resolved joint \nicer and \nustar spectra fitting with the absorbed Comptonization model
(see text for detail). Phase intervals values are given in the each panel. 
}
\label{fig:ph_spc_evol}
\end{figure}

\subsection{Pulse phase-resolved spectroscopy}
It is well established that spectra of XRPs vary
with the pulse phase. Parameters of the cyclotron resonant scattering features (CRSFs)
are also know to change at such time scales
\citep[see, e.g.,][and references therein]{2000ApJ...530..429B,2004A&A...427..975K,2004AIPC..714..323H,2015MNRAS.448.2175L},
and, in some cases, lines can only appear significantly at certain phase
intervals \citep{2019ApJ...883L..11M}.
Therefore the pulse phase-resolved spectroscopy can be considered as a tool for
the study of the line properties, and ultimately for probing the geometry of
the emission regions in the vicinity of the neutron star and its magnetic field structure.
Here we focus on understanding whether the absorption features are detected at
individual pulse phases in order to exclude the situation when the detection of
the features in the averaged spectrum arises from the modeling of superimposed
spectra variable across different pulse phases.

As a first step, we fitted phase-resolved \nustar and \nicer spectra extracted
from the 0.2 phase length intervals with the Comptonization model continuum modified
by the interstellar absorption but without absorption or emission like features and
soft black-body component.
Residuals of the fits relative to the absorption lines (Fig.\,\ref{fig:ph_spc_evol}) 
are observed throughout the pulse, although the depth of the individual features
appears to be variable and is most clearly seen at the phase interval 0.65-0.85.

\begin{figure}
\vspace{-2cm}
\hspace{-0.4cm}\includegraphics[width=1.1\columnwidth]{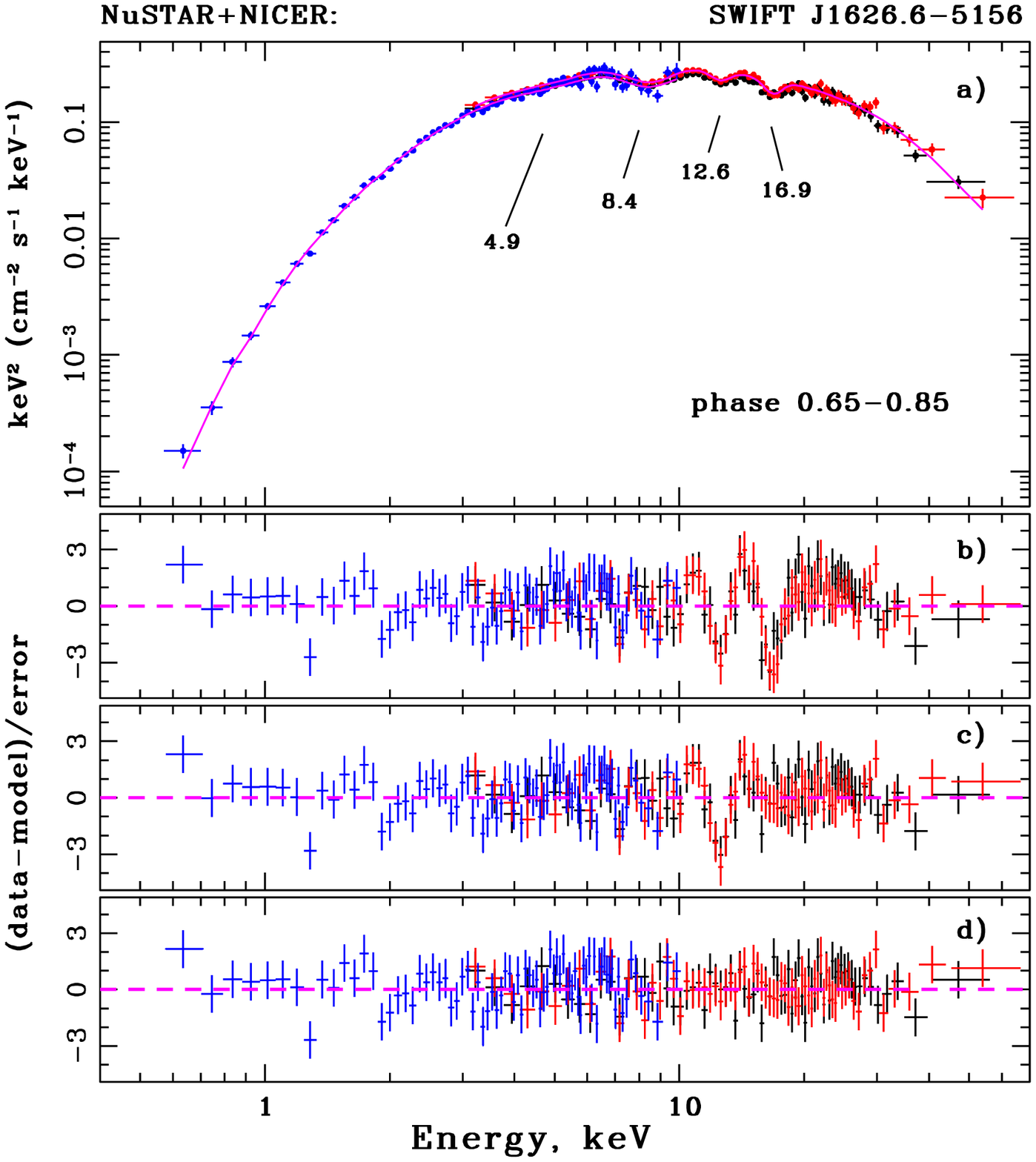}
\vspace{-2.5cm}
\caption{(a) The energy spectrum of \src at the pulse phases 0.65-0.85 reconstructed
with {\it NuSTAR} and \nicer and fitted with the Comptonization model modified by interstellar absorption
and four absorption features (see the text for details). Residuals for spectral model: without two absorption
lines with the highest energies (b), including line near 17 keV (c) and including all lines (d).} 
\label{fig:ph_cycl_spec}
\end{figure}

On the next step, we fitted these spectra adding up to four absorption-like features.
Unlike to the phase-averaged analysis, we did not include the soft black body
component (fits are not sensitive to that, probably due to the lack of statistics
in phase-resolved spectra) and the iron line. 
At all phases, the inclusion the two absorption features at $E\sim 5$ keV and $E\sim9$ keV
is necessary to obtain significant fits. For phases $0.05-0.25$ and $0.25-0.45$ these two
components are actually sufficient. For phases $0.45-0.65$ and $0.85-1.05$ the inclusion
of an additional absorption line at $E\sim17$ keV strongly improves the fit quality.
In first case, the value of the $\chi ^2$ changes from 730 (650 dof) to 681 (647 dof), and
in the second, from 671 (579 dof) to 644 (576 dof). In terms of Bayesian criteria, we
obtain $\Delta BIC=41$ and $\Delta BIC=19$ for the first and the second cases, respectively,
which implies (since both values are $>10$) that the strength of statement that the model
with three lines is better than with two lines is "Very strong". Finally, to achieve an
acceptable fit for the spectrum at phase $0.65-0.85$, all four absorption lines have to
be included. More specifically, the $\chi^2$ value changes from 796 (660 dof) to 704 (657 dof)
after adding the line at $\sim17$ keV, and reduces to 630 (564 dof) after a fourth absorption
line feature at $\sim13$ keV is included (see Fig.,\ref{fig:ph_cycl_spec}). Bayesian
information criterion decreases on 83 and 65, respectively. Results of the approximation
of the phase-resolved spectra are summarized in Table\,\ref{tab:spe_phase}.

Thus we have confirmed presence of all four absorption features in
the phase-resolved spectra as well. However, not all lines are detected in all phase bins.
We also note that their energies vary slightly with the pulse phase. In addition, we note
that to describe the phase-resolved spectra, it is not necessary to include a component
for the iron line. 

\begin{table*}[!h]
\small
\begin{center}
\caption{Best-parameters of the \src phase-resolved spectra fitting with the Comptonization model
modified by interstellar absorption and with inclusion of up to four cyclotron-line absorption features}
\label{tab:spe_phase}
\begin{tabular}{l|c|c|c|c|c}
\hline
\hline
$phase$ & {\small\bf 0.05-0.25} & {\small\bf 0.25-0.45} & {\small\bf 0.45-0.65} & {\small\bf 0.65-0.85} & {\small\bf 0.85-1.05}\\
\hline
Parameter  & Value  & Value & Value  & Value & Value\\
\hline
$N_{\rm H}$                &  $0.37\pm0.02$  & $0.44\pm0.03$  &  $0.43\pm0.02$  &  $0.42\pm0.02$  & $0.28\pm0.03$   \\ 
\hline
$T_0$, keV                 &  $1.14\pm0.03$  & $1.10\pm0.03$  &  $1.17\pm0.03$  &  $1.18\pm0.03$  &  $1.21\pm0.08$  \\
$T_{\rm p}$, keV           &  $6.38\pm0.37$  & $6.25\pm0.25$  &  $6.24\pm0.28$  &  $6.06\pm0.21$  &  $6.57\pm0.58$  \\
$\tau_{\rm p}$             &  $2.86\pm0.23$  & $3.19\pm0.20$  &  $3.13\pm0.21$  &  $3.38\pm0.19$  &  $2.85\pm0.40$  \\
$A_{\rm comp}\times10^{1}$ &  $0.114\pm0.002$& $0.142\pm0.005$&  $0.180\pm0.007$&  $0.191\pm0.006$&  $0.103\pm0.007$\\
\hline
$E_{cyc1}$, keV            &  $4.88\pm0.06$  & $4.99\pm0.05$  &  $5.00\pm0.06$  &  $4.90\pm0.06$  &  $4.61\pm0.08$  \\
$\sigma_{cyc1}$, keV       &  $0.75\pm0.10$  & $0.72\pm0.08$  &  $0.95\pm0.07$  &  $1.02\pm0.10$  &  $0.87\pm0.11$  \\
$\tau_{cyc1}$, keV         &  $0.30\pm0.08$  & $0.33\pm0.08$  &  $0.53\pm0.08$  &  $0.47\pm0.12$  &  $0.42\pm0.12$  \\
\hline
$E_{cyc2}$, keV            &  $8.79\pm0.08$  & $9.22\pm0.12$  &  $8.54\pm0.05$  &  $8.39\pm0.04$  &  $9.02\pm0.16$  \\
$\sigma_{cyc2}$, keV       &  $1.46\pm0.16$  & $1.73\pm0.19$  &  $1.41\pm0.09$  &  $1.09\pm0.09$  &  $2.19\pm0.42$  \\
$\tau_{cyc2}$, keV         &  $0.93\pm0.21$  & $0.98\pm0.24$  &  $1.20\pm0.15$  &  $0.91\pm0.17$  &  $1.64\pm0.90$  \\
\hline
$E_{cyc3}$, keV            &  $-$            & $-$            &  $-$            &  $12.56\pm0.08$ &  $-$            \\
$\sigma_{cyc3}$, keV       &  $-$            & $-$            &  $-$            &  $0.64\pm0.13$  &  $-$            \\
$\tau_{cyc3}$, keV         &  $-$            & $-$            &  $-$            &  $0.28\pm0.07$  &  $-$            \\
\hline
$E_{cyc4}$, keV            &  $-$            & $-$            &  $17.55\pm0.22$ &  $16.86\pm0.09$ &  $16.90\pm0.26$ \\
$\sigma_{cyc4}$, keV       &  $-$            & $-$            &  $1.34\pm0.26$  &  $0.84\pm0.11$  &  $1.23\pm0.35$  \\
$\tau_{cyc4}$, keV         &  $-$            & $-$            &  $0.56\pm0.13$  &  $0.58\pm0.08$  &  $0.54\pm0.21$  \\
\hline
$C_{\rm modB}$             &  $1.063\pm0.006$& $1.067\pm0.006$&  $1.075\pm0.005$&  $1.075\pm0.005$&  $1.074\pm0.007$\\ 
$C_{\rm NIC}$            &  $0.933\pm0.011$& $0.960\pm0.011$&  $0.979\pm0.010$&  $0.976\pm0.010$&  $0.917\pm0.012$\\ 
$F_{X}\times10^{10}$   &  $3.6$          & $4.5$          &  $5.7$          &  $6.1$          &  $3.1$          \\ 
\hline
$\chi^2$ (d.o.f.)          &  601.3 (592)    & 745.6 (626)    &  681.2 (647)    &  630.1 (654)    &  643.8 (576)    \\
\hline
\end{tabular}
\end{center}
\begin{tabular}{l l}
\small
{\bf Notes.} & See Notes to Table~\ref{tab:spe_mean} \\
\end{tabular}

\end{table*}

\section{Discussion and conclusions}

Most of the known cyclotron line sources exhibit either only the fundamental CRSF or the fundamental one
and its first harmonic. This is likely a selection effect associated with the difficulty of detecting such
features at higher energies. In fact, most accreting pulsars are strongly magnetized with fundamental line
energies typically above the a cutoff energy at around 20\,keV (see e.g., \citealt{2019A&A...622A..61S}
for a recent review). The detection of the first and certainly of the second harmonics is challenging due
to the lack of photons well above the cutoff.  

In our analysis, we not only confirm the CRSFs at $8.6$ and $17.1$ keV in the spectrum of \src already
reported in literature, but also discover two additional features around $\sim 4.9$ keV and $\sim13$\,keV.
We conclude, therefore, that four CRSFs characterize the spectrum of \src, with the fundamental line
at $E\sim4.9$\,keV, and other three features being harmonics of this feature. This implies that \src has
the lowest confirmed magnetic field among all X-ray pulsars, and is only second to 4U~0115+63
\citep{1999ApJ...523L..85S} by total number of observed cyclotron lines. 

In the case of \src the detection of four lines is only possible due to the low energy of the
fundamental one. The strength of the neutron star magnetic field is thus estimated
to be $B\sim4.1(1+z)\times10^{11}$\,G. The only other XRP with the comparable field is the
peculiar ``bursting pulsar'' GRO~J1744-28 \citep{2015MNRAS.452.2490D,2015MNRAS.449.4288D}.
Our result is rather unexpected and it is interesting to compare the magnetic field
strength estimated through the CRSF with other indirect estimates.

\begin{figure}[!t]
\vspace{-2cm}
\includegraphics[width=1.\columnwidth]{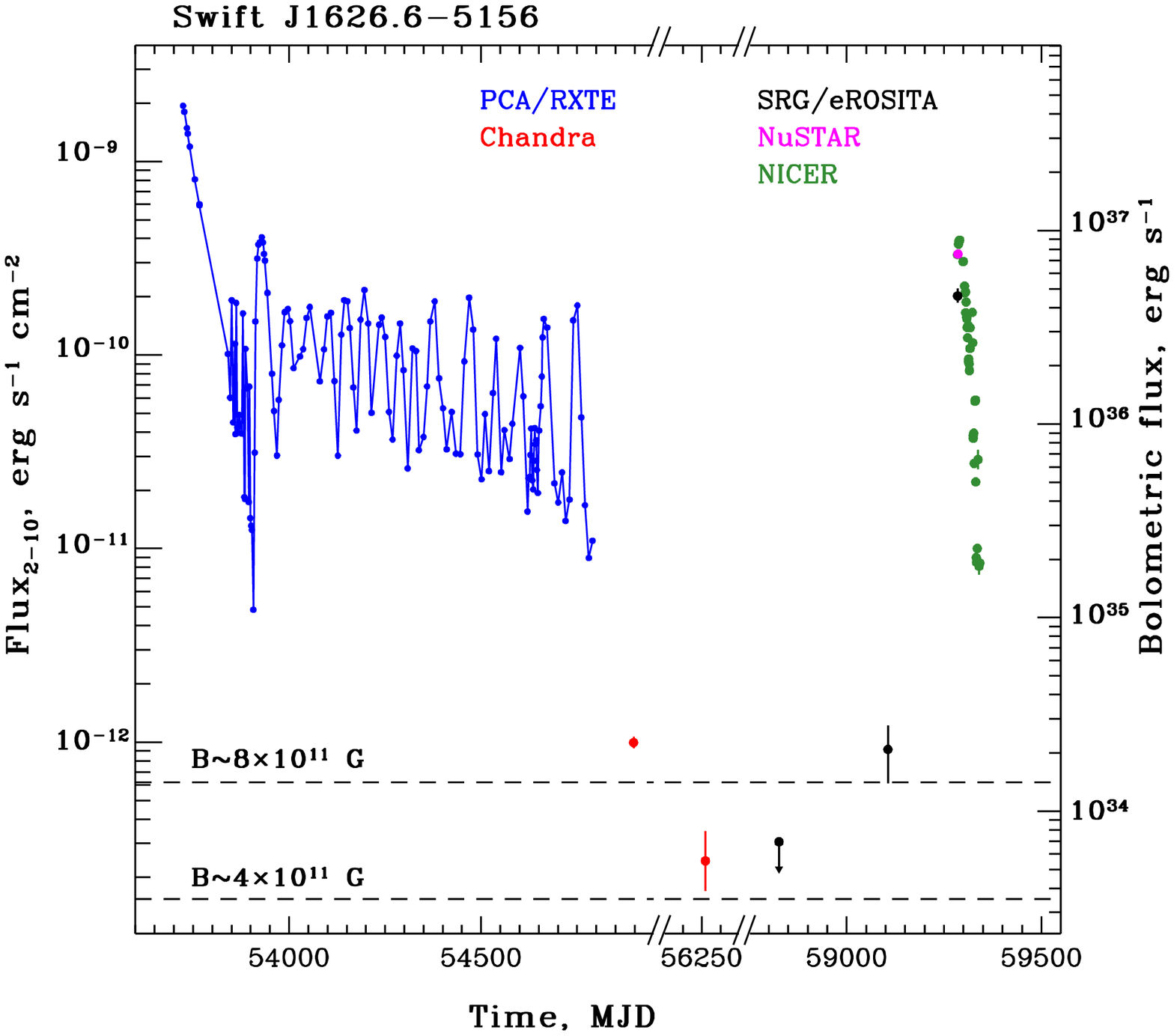}
\vspace{-3cm}
\caption{\src long-term flux history. Bolometric luminosity calculated in assumption of 10 kpc distance
to the source. Dashed lines show luminosity when the source should drop to the propeller regime
for two magnetic field values corresponding to 4.9 and 9 keV fundamental cyclotron lines.}
\label{fig:lc_longterm}
\end{figure}

First, as already mentioned, we note that the source continues to accrete in
quiescence \citep{2017MNRAS.470..126T} down
to a bolometric luminosity of $\simeq5.9\times10^{33} d_{10}^2$\,erg\,s$^{-1}$ (here we recalculated
the lowest observed luminosity in the 0.5-10 keV energy band to the bolometric one based on our
knowledge of the source broad band spectrum and $d_{10}$ is distance to the source scaled to 10\,kpc).
This is consistent with the observed \ero\ flux and the source flux variability at low luminosities
as illustrated by Fig.~\ref{fig:lc_longterm}. If the accretion continues, then the lowest luminosity
value must have been
higher than the limiting luminosity for the transition to the propeller
regime \citep{2016A&A...593A..16T}:
$$
L_{lim}=4\times10^{37}k^{7/2}B_{12}^2 P^{-7/3} M_{1.4}^{-2/3} R_6^5 \le L_{10} d_{10}^2
$$
here $k$ is a factor relating to the size of the magnetosphere for a given accretion configuration
to the Alfven radius and $B_{12}$ is the magnetic field in units
of $10^{12}$\,G and $L_{10}$ is a
lowest measure bolometric flux of the source calculated in assumption of
10 kpc distance. Assuming canonical neutron star
parameters {of $M=1.4 M_{\sun}$, $R=10^6$~cm and $k=0.5$},
the measured source minimal luminosity $L_{10}=5.9\times 10^{33}$ erg/s and the spin period of 15.4 s one
obtains the magnetic
field value $B_{12}\le 1 \times d_{10}$.
Considering the limits on distance from Gaia EDR3 of 5.8-12\,kpc
\citep{2021AJ....161..147B}, $B_{12}\le(0.58-1.2)\times10^{12}$\,G. Although formally consistent
with interpretation of either 4.9\,keV and 9\,keV lines as fundamental, the line at 9\,keV is
already at the edge of the allowed distance range and inconsistent with the best photo-geometric
Gaia estimate of 6.6\,kpc. We conclude, therefore, that also observed properties of \src in quiescence
favor the low field implied by $\sim4.9$\,keV fundamental cyclotron line energy.

This makes \src the weakest magnetised classical X-ray pulsar among all cyclotron line sources and
might have relevant consequences for the science program of upcoming X-ray polarimeters such
as {\it IXPE} \citep{2016SPIE.9905E..17W}, and the polarisation focusing array (PFA) on board
the \textit{eXTP} mission \citep{2016SPIE.9905E..1QZ,2019SCPMA..6229505S}.
Indeed, those instruments operate in the soft X-ray band where the emission of XRPs is expected to be
strongly polarized due to the birefringence induced by a Compton scattering in the strong magnetic
field \citep{1988ApJ...324.1056M} for ordinary and extraordinary mode photons. The cross-sections for
photons with both modes become comparable around the cyclotron resonance energy, which makes this energy
extremely interesting. For the vast majority of XRPs the cyclotron resonance energy lies outside of the
operational range of the gas pixel detectors used in \textit{IXPE}/\textit{eXTP} polarimeters. The only
exceptions are \src and GRO~J1744-28. However GRO~J1744-28 has a very low outburst duty cycle and thus
it is unlikely that will be observed. On the other hand, \src appears to be regularly detectable by all
sky monitors and wide field X-ray instruments and thus is the ideal target for the soon to come
X-ray polarimeters, especially if it undergoes another giant outburst.

\section*{Acknowledgements}
We thank the {\it NuSTAR} team for the help with organising prompt observation.
This work was financially supported by the Russian Science Foundation (grant 19-12-00423).



\bibliography{swiftj1626.6}
\bibliographystyle{aasjournal}
\label{lastpage}
\end{document}